\documentstyle[aps,twocolumn,prb,epsf]{revtex}
\begin{document}
\title{Estimation of the charge carrier localization length from Gaussian
fluctuations in the magneto-thermopower of $La_{0.6}Y_{0.1}Ca_{0.3}MnO_3$}
\author{S. Sergeenkov$^{1,2}$, H. Bougrine$^{1,3}$, M. Ausloos$^{1}$, and
A. Gilabert$^{4}$}
\address{$^{1}$SUPRAS, Institute of Physics, B5, University of Li$\grave e$ge,
B-4000 Li$\grave e$ge, Belgium\\
$^{2}$Bogoliubov Laboratory of Theoretical Physics,
Joint Institute for Nuclear Research, 141980 Dubna, Moscow Region, Russia\\
$^{3}$SUPRAS, Montefiore Electricity Institute, B28, University of
Li$\grave e$ge, B-4000 Li$\grave e$ge, Belgium\\
$^{4}$Laboratoire de Physique de la Mati$\grave e$re Condens\'ee, Universit\'e
de Nice-Sophia Antipolis,\\ Parc Valrose F-09016 Nice, Cedex 02, France\\}
\date{\today}
%\preprint
\draft
\maketitle
\begin{abstract}
The magneto-thermoelectric power (TEP) $\Delta S(T,H)$ of perovskite type
manganise oxide $La_{0.6}Y_{0.1}Ca_{0.3}MnO_3$ is found to exhibit a sharp
peak at some temperature $T^{*}=170K$.
By approximating the true shape of the measured
magneto-TEP in the vicinity of $T^{*}$ by a linear
triangle of the form $\Delta S(T,H)\simeq S_p(H)\pm B^{\pm}(H)(T^{*}-T)$,
we observe that $B ^{-}(H)\simeq 2B ^{+}(H)$.
We adopt the electron localization scenario and introduce
a Ginzburg-Landau (GL) type theory which
incorporates the two concurrent phase transitions, viz., the
paramagnetic-ferromagnetic transition at the
Curie point $T_C$ and the "metal-insulator" (M-I) transition at
$T_{MI}$. The latter is characterized by the divergence of the
field-dependent charge carrier localization length $\xi (T,H)$ at some
characteristic field $H_0$.
Calculating the average and fluctuation contributions to the total
magnetization and the transport entropy related magneto-TEP
$\Delta S(T,H)$ within the GL theory, we obtain a simple relationship
between $T^{*}$ and the above two critical temperatures ($T_{C}$ and
$T_{MI}$). The observed slope ratio $B ^{-}(H)/B ^{+}(H)$ is found to be
governed
by the competition between the electron-spin exchange $JS$ and the
induced magnetic energy $M_sH_0$. The comparison of our data with
the model predictions produce $T_{C}=195K$, $JS=40meV$, $M_0=0.4M_s$,
$\xi _0=10\AA$, and $n_e/n_i=2/3$ for the estimates of the Curie temperature,
the exchange coupling constant, the critical magnetization, the localization
length, and the free-to-localized carrier number density ratio, respectively.
\end{abstract}

\narrowtext

\section{Introduction}

The intriguing magnetotransport properties of manganite's family
$R_{1-x}A_{x}MnO_3$ (where $R=La,Y,Nd,Pr$ and $A=Ca,Sr,Ba,Pb$)
with a $Mn^{3+}/Mn^{4+}$ mixed valence keep attracting much attention of
both experimentalists and theorists.~\cite{1,2,3,4,5,6,7,8,9,10,11,12,13,14}
In the doping range $0.2<x<0.5$, these compounds are known to undergo a double
phase transition from paramagnetic (PM) insulator (I) to ferromagnetic (FM)
metal (M) state
characterized by the Curie temperature $T_C$ and the charge carrier
localization
temperature $T_{MI}$, respectively. The so-called giant magnetoresistivity
(GMR) exhibits a sharp peak around $T_{MI}$, while below $T_C$ the system
acquires a spontaneous magnetization accompanied by a giant
magnetic entropy changes.~\cite{14}
Despite a
variety of theoretical scenarios attempting to describe this phenomenon,
practically all of them adopt as a starting point the so-called
double-exchange (DE) mechanism, which considers the exchange of electrons
between neighboring $Mn^{3+}/Mn^{4+}$ sites with strong on-site Hund's
coupling. The estimated exchange energy~\cite{11} $JS=45meV$ (where $S=2$
is an effective
spin on a $Mn$ site), being much less than the Fermi energy $E_F$ in these
materials (typically, $E_F=0.15eV$), favors an FM ground state. In turn,
an applied magnetic field $H$ enhances the FM order thus reducing
the spin scattering and producing the observed negative GMR.
The localization scenario,~\cite{13} in which $Mn$ oxides are modelled
as systems with both DE off-diagonal spin disorder and nonmagnetic diagonal
disorder,
predicts a divergence of the electronic localization length $\xi (M)$
at some M-I phase transition.
In terms of the spontaneous magnetization $M$,
it means that for $M<M_0$ the system is in a highly resistive (insulator-like)
phase, while for $M>M_0$ the system is in a low resistive (metallic-like)
state.
Within this scenario, the Curie point $T_C$
is defined through the spontaneous magnetization $M$ as $M(T_C,H)=0$,
while the M-I transition temperature $T_{MI}$ is such that $M(T_{MI},H)=M_0$
(with $M_0$ being a fraction of the saturated magnetization $M_s$).
Furthermore, the influence of magnetic fluctuations on electron-spin
scattering near $T_{MI}$ is expected to be rather important, for they can
easily tip a subtle balance between magnetic and electronic processes in
favor of either charge localization or delocalization.
Besides, the observable
difference between the two critical temperatures (usually attributed
to the quality of a particular sample used~\cite{5,6,7,8}) is ascribed to
the random nonmagnetic scattering which is highly responsible
for the magnitude of the observable GMR.~\cite{13}

On the other hand, in view of its carrier charge (and density) sensitive
nature, thermopower (TEP) measurements could complement the traditional
MR data and be used as a tool for probing the field-induced delocalization
of the carriers.
Indeed, studying the observable magneto-TEP
$\Delta S(T,H)=S(T,H)-S(T,0)$ has already proved to be useful for providing
important insights into different aspects of high-$T_c$ superconductors
in the mixed state.~\cite{15,16,17} Besides, magneto-TEP can be directly
linked to the transport entropy change in applied magnetic field.
The recently observed~\cite{14} giant magnetic entropy change in
manganites (produced by the abrupt reduction of the magnetization
and attributed to an anomalous thermal expansion just at the Curie point)
gives another reason to utilize the magneto-TEP data in order to get
an additional information as for the underlying transport mechanisms in
these materials.

In the present paper we discuss some typical results
for magneto-TEP measurements on a manganite sample
$La_{0.6}Y_{0.1}Ca_{0.3}MnO_3$ at $H=1T$ field for a wide temperature
interval (ranging from $20K$ to $300K$). By approximating the true shape of
the measured magneto-TEP in the vicinity of the peak temperature $T^{*}$
by a linear triangle of the form $\Delta S(T,H)\simeq S_p(H)\pm
B^{\pm}(H)(T^{*}-T)$, we observe that $B ^{-}(H)\simeq 2B ^{+}(H)$.
In an attempt to account for the
observed behavior of the magneto-TEP, we adopt the main ideas of the
microscopic localization theory~\cite{13} and construct
a phenomenological free energy functional of Ginzburg-Landau (GL) type
which describes the magnetic field and temperature behavior of the spontaneous
magnetization in the presence of strong localization effects near $T^{*}$.
Calculating the background and fluctuation contributions to the total
magnetization and the transport entropy-induced magneto-TEP
$\Delta S(T,H)$ within the GL theory, we obtain a simple relationship
between $T^{*}$ and the above two critical temperatures ($T_{C}$ and $T_{MI}$).
We find also that the observed ratio $B ^{-}(H)/B ^{+}(H)$ asymmetry is governed
by a universal parameter $z=JS/M_sH_0$ where $JS$ is
the electron-spin exchange and $M_sH_0$ is the localization related magnetic
energy.
By comparing our data with the model predictions, we deduce estimates for
some important model parameters such as the Curie point $T_{C}$, the
localization length $\xi _0$, the critical magnetization $M_0\propto H_0$,
the exchange energy $J$, and the free-to-localized carrier number density
ratio $n_e/n_i$, all in good agreement with the existing microscopic
localization theories.

\section{Experimental results}

$La_{0.6}Y_{0.1}Ca_{0.3}MnO_3$ samples were prepared from
stoichiometric
amounts of $La_{2}O_3$, $Y_{2}O_3$, $CaCO_3$, and $MnO_2$ powders.
The mixture was heated in the air at $800C$ for 12 hours to achieve
the decarbonation. Then it was pressed at room temperature under
$10^3kG/cm^2$ to obtain parallelipedic pellets. An annealing and sintering
from $1350C$ to $800C$ was made slowly (during 2 days) to preserve the
right phase stoichiometry. A small bar (length $l=10mm$, cross section
${\cal S}=4mm^2$)
was cut from one pellet. The electrical resistivity $\rho (T,H)$ was
measured using the conventional four-probe method. To avoid Joule and
Peltier effects, a dc current $I=1mA$ was injected (as a one second pulse)
successively on both sides of the sample. The voltage drop $V$ across the
sample was measured with high accuracy by a $KT256$ nanovoltmeter. The
magnetic field $H$ of $1T$ was applied normally
to the current.
Fig.1 presents the temperature dependence of the magnetoresistance (MR)
$\Delta \rho (T,H)=\rho (T,H)-\rho (T,0)$ for a $La_{0.6}Y_{0.1}Ca_{0.3}MnO_3$
sample at $H=1T$ field. As is seen, the negative MR $\Delta \rho (T,H)$
shows a peak (dip) at some temperature $T_{0}=160K$ (referred to as $T_{MI}$,
in what follows) where the GMR $\Delta \rho (T,H)/\rho (T,0)$ reaches $40\%$.
The thermopower (TEP) $S$ was measured using the differential
method.~\cite{18} In order to generate a heat flow, a small heater film
($R=150\Omega$) was attached to one end of the sample. Two calibrated
chromel-constantan thermocouples were used to measure the temperature
difference between two points on the sample. The TEP $S(T,H)$ is
deduced from the following equation, $S(T,H)=S_{Au}(T)-V_s(T,H)/\Delta T$,
where $S_{Au}(T)$ is the TEP of the gold wires used to measure the voltage
drop $V_s$ at the hot junctions of both thermocouples.
Fig.2 shows a typical temperature behavior of the deduced magneto-TEP
$\Delta S(T,H)=S(T,H)-S(T,0)$ for the same sample (at $H=1T$).
Observe that it has an asymmetric $\Lambda$-like shape near
some critical temperature $T^{*}>T_{MI}$ where it reaches its
field-dependent peak (dip) value $S_p(H)$.
Approximating the shape of the observed $\Delta S(T,H)$ by the asymmetric
linear triangle of the form
\begin{equation}
\Delta S(T,H)\simeq S_p(H)\pm B^{\pm}(H)(T^{*}-T),
\end{equation}
with positive slopes $B ^{-}(H)$ and $B ^{+}(H)$ defined for $T<T^{*}$ and
$T>T^{*}$, respectively, we find (see Fig.2) that $B^{-}(H)\simeq 2B^{+}(H)$
in the vicinity of $T^{*}$.
Now, with all this information in mind, let us proceed to the interpretation
of the experimental results.

\section{Discussion}

\subsection{The model}

Since we are dealing with the magnetic-field induced changes of the TEP,
it is reasonable to
assume that the observed behavior can be attributed to the corresponding
changes of transport magnetic entropy (and thus spontaneous magnetization)
in the presence of strong electron-spin exchange and localization
effects, near some critical temperature $T^{*}$. Later on, we will establish
a simple (linear) relationship between the peak temperature $T^{*}$ and
the two critical temperatures $T_C$ and $T_{MI}$, responsible respectively
for PM-FM and M-I phase transitions.
Based on the above considerations, we can write
${\cal F}={\cal F}_M-{\cal F}_e$
for the balance of magnetic ${\cal F}_M$ and electronic ${\cal F}_e$
free energies participating in the transport processes under discussion.
The observed magnetization $M$ and the magneto-TEP
behavior should result from the minimization of ${\cal F}$ (as, for example,
is the case in superconductors where ${\cal F}$ measures the difference
between the normal and condensate energies~\cite{15,16}). In our case,
the above electronic contribution reads
${\cal F}_e={\cal M}{\cal H}_e=\eta ^2(n_eE_k+n_iV_{DE})$ and
describes a coupling of spontaneous magnetization ${\cal M}=M_s\eta ^2$
(where $\eta $ is the order parameter and $M_s$ the saturated magnetization)
with (i) an effective DE energy $V_{DE}=-JS$ (where $S$ is an effective spin
on a $Mn$ site, and $J$ the exchange coupling constant),
and (ii) the electronic (localization) energy $E_k(T,H)=\hbar ^2/2m\xi ^2(T,H)$
(where $\xi (T,H)$ is the localization length, and $m$ an effective electron
mass); $n_i$ and $n_e$ stand for the number density of localized spins and
conduction electrons, respectively. At the same time, the magnetic
contribution
${\cal F}_M={\cal M}({\cal H}_{eff}-H)=M_s\eta ^2 (\gamma \eta ^2-H)$
includes the effects due to the molecular-field ${\cal H}_{eff}=
\gamma {\cal M}/M_s$ (where
$\gamma =3k_BT_C/2\mu _BS$ is the characteristic magnetic field with $k_B$
the Boltzman constant and $\mu _B$ the Bohr magneton) and an applied magnetic
field $H$. After trivial rearrangements, the above functional ${\cal F}$ can
be cast into a familiar GL type form describing the second-order phase
transition from PM (insulator) to FM (metal) state near $T^{*}$, namely
\begin{equation}
{\cal F} [\eta ]=a\eta ^2 +\frac{\beta}{2}\eta ^4-\zeta \eta ^2.
\end{equation}
Here $\zeta (H)=M_sH-n_iJS$ is the
effective field-dependent chemical potential of quasiparticles;
$a(T,H)=\alpha (H)(T-T^{*})$ with $\alpha (H)=n_e\hbar ^2/2m\xi ^2_0(H)T^{*}$;
$\beta =2\gamma M_s$, and we used the conventional expression
$\xi ^2(T,H)=\xi ^2_0(H)/(1-T/T^{*})$ for the correlation length. Besides,
to account for the field-induced localization effects, we assume after Sheng
et al.~\cite{13} that $\xi _0(H)/\xi _0(0)=1/(1-H/H_0)$
with $H_0\simeq \gamma \propto M_0$.

\subsection{Mean value of the magneto-TEP: $\Delta S_{av}(T,H)$}

Given our previous experience with high-$T_c$ superconductors, we
can readily present the observed magneto-TEP in a two-term contribution
form~\cite{16}
\begin{equation}
\Delta S(T,H)=\Delta S_{av}(T,H)+\Delta S_{fl}(T,H),
\end{equation}
where the average term $\Delta S_{av}(T,H)$ is non-zero only
below $T^{*}$ while the fluctuation term $\Delta S_{fl}(T,H)$
should contribute to the observable $\Delta S(T,H)$ both above and below
$T^{*}$. In what follows, we shall discuss these two contributions separately
within a mean-field theory approximation for GMR materials.

As usual, the equilibrium state of such a system is determined from the
minimum energy condition $\partial {\cal F}/\partial \eta =0$ which
yields for $T<T^{*}$
\begin{equation}
\eta _0^2=\frac{\alpha (H)(T^{*}-T)+\zeta (H)}{\beta }
\end{equation}
Substituting $\eta _0$ into Eq.(2) we obtain for the average free
energy density
\begin{equation}
\Omega _{av}(T,H)\equiv {\cal F} [\eta _0]=-
\frac{[\alpha (H)(T^{*}-T)+\zeta (H)]^2}{2\beta }
\end{equation}
In turn, the magneto-TEP $\Delta S(T,H)$ can be related to the
corresponding difference of transport entropies~\cite{15,16,17}
$\Delta \sigma _{av}\equiv
-\partial \Delta \Omega _{av}/\partial T$ as $\Delta S_{av}(T,H)=
\Delta \sigma _{av}(T,H)/en_e$,
where $e$ and $n_e$ are the charge and the number density of free carriers.
Finally the mean value of the magneto-TEP reads (below $T^{*}$)
\begin{equation}
\Delta S_{av}(T,H)=S_{p,av}(H)-B _{av}(H)(T^{*}-T),
\end{equation}
with
\begin{equation}
S_{p,av}(H)=-\frac{\alpha (0)\Delta \zeta (H)}{e\beta n_e}(1+z),
\end{equation}
and
\begin{equation}
B _{av}(H)=\frac{2\alpha (0)\Delta \alpha (H)}{e\beta n_e},
\end{equation}
where $z=\Delta \alpha (H)\zeta (0)/\alpha (0)\Delta \zeta (H)$ with
$\Delta \alpha (H)=\alpha (H)-\alpha (0)$ and $\Delta \zeta (H)=\zeta (H)-
\zeta (0)$.

\subsection{Mean-field Gaussian fluctuations of the magneto-TEP:
$\Delta S_{fl}(T,H)$}

The influence of fluctuations (both Gaussian and critical) on transport
properties of high-$T_c$ superconductors
(including TEP, electrical and thermal conductivity) was extensively
studied and is very well documented
(see, e.g.,~\cite{19,20,21,22,23,24,25} and further
references therein). In particular, it was found that the
fluctuation-induced behavior may extend to temperatures more than
$10K$ higher than the critical temperature $T_c$. As for manganites,
the fluctuation effects in these materials appear to be much less
explored.~\cite{26} Nonetheless, according to the
interpretation of the observed magneto-TEP we adopt in the
present paper, influence of magnetic fluctuations on electron-spin
scattering near $T^{*}$ should be rather important.
So, it seems appropriate to take a closer look at the region near
$T^{*}$ to discuss the fluctuations of the magneto-TEP
$\Delta S_{fl}(T,H)$.
Recall that according to the textbook theory of Gaussian
fluctuations,~\cite{27}
the fluctuations of any observable (such as heat capacity, magnetization,
etc) which is conjugated to the order parameter $\eta$ can be presented
in terms of the statistical
average of the fluctuation amplitude $<(\delta \eta )^2>$ with
$\delta \eta =\eta -\eta _0$. Then the TEP above $(+)$ and
below $(-)$ the critical point $T^{*}$ have the form of
\begin{equation}
S_{fl}^{\pm}(T,H)=A<(\delta \eta )^2>_{\pm}
=\frac{A}{Z}\int d\eta (\delta \eta )^2 e^{-\Sigma [\eta ]},
\end{equation}
where
$Z=\int d\eta e^{-\Sigma [\eta ]}$ is the partition function with
$\Sigma [\eta ]\equiv ({\cal F} [\eta ]-{\cal F} [\eta _0])/k_BT$, and $A$
is a coefficient to be defined below.
Expanding the free energy density functional ${\cal F} [\eta ]$
\begin{equation}
{\cal F} [\eta ]\approx {\cal F} [\eta _0]+
\frac{1}{2}\left[ \frac{\partial ^2{\cal F}}
{\partial \eta ^2}\right ]_{\eta =\eta _0}\!(\delta \eta )^2,
\end{equation}
around the mean value of the order parameter $\eta _0$, which is defined as a
stable solution of equation $\partial {\cal F}/\partial \eta =0$ we can
explicitly calculate the Gaussian integrals.
Due to the fact that $\eta _0$ is given by Eq.(4) below $T^{*}$ and
vanishes at $T\ge T^{*}$, we obtain finally
\begin{equation}
S_{fl}^{-}(T,H)=\frac{Ak_BT^{*}}{4\alpha (H)(T^{*}-T)+4\zeta (H)},
\qquad T\le T^{*}
\end{equation}
and
\begin{equation}
S_{fl}^{+}(T,H)=\frac{Ak_BT^{*}}{2\alpha (H)(T-T^{*})-2\zeta (H)},
\qquad T\ge T^{*}
\end{equation}
As we shall see below, for the experimental range of parameters under
discussion, $\zeta (H)/\alpha (H)\gg |T^{*}-T|$. Hence, with a good
accuracy we can linearize Eqs.(11) and (12) and obtain for the fluctuation
contribution to the magneto-TEP
\begin{equation}
\Delta S_{fl}^{\pm}(T,H)\simeq S_{p,fl}^{\pm}(H)\pm B _{fl}^{\pm}(H)(T^{*}-T),
\end{equation}
where
\begin{equation}
S_{p,fl}^{-}(H)=-\frac{1}{2}S_{p,fl}^{+}(H)=
-\frac{Ak_BT^{*}\Delta \zeta (H)}{4\zeta ^2(0)},
\end{equation}
and
\begin{equation}
B _{fl}^{-}(H)=-\frac{1}{2}B _{fl}^{+}(H)=
-\frac{Ak_BT^{*}\Delta \alpha (H)}{4\zeta ^2(0)}\left(
1-\frac{2}{z}\right).
\end{equation}
Furthermore, it is
quite reasonable to assume that $S_p^{-}=S_p^{+}\equiv
S_p$, where the magneto-TEP peak (dip) values are defined as follows,
$S_{p}^{-}=S_{p,av}+S_{p,fl}^{-}$ and $S_{p}^{+}=S_{p,fl}^{+}$.
The above equations allow us to fix the arbitrary parameter $A$
yielding $A=-4\zeta ^2(0)\alpha (0)(1+z)/3ek_BT^{*}\beta n_e$.
This in turn leads
to the following expressions for the fluctuation contribution to
peaks and slopes through their average counterparts (see Eqs.(7) and (8)):
$S_{p,fl}^{+}(H)=(2/3)S_{p,av}(H)$, $S_{p,fl}^{-}(H)=-(1/3)S_{p,av}(H)$,
$B _{fl}^{-}(H)=-(1/2)B_{av}(H)$, and $B _{fl}^{+}(H)=B_{av}(H)$.
Finally, the total contribution to the observable magneto-TEP reads
(Cf. Eq.(1))
\begin{equation}
\Delta S(T,H)=S_p(H)\pm B^{\pm}(H)(T^{*}-T),
\end{equation}
where
\begin{equation}
S_{p}(H)=-\frac{(1+z)E_k^0}{3eT^{*}}\left(\frac{H}{H_0}\right),
\end{equation}
\begin{equation}
B^{+}(H)\equiv B _{fl}^{+}(H)=\left(\frac{n_e}{n_i}\right)
\frac{(z-2)E_k^0}{JST^{*}}S_{p}(H),
\end{equation}
and
\begin{eqnarray}
B^{-}(H)&\equiv &B_{av}(H)+B^{-}_{fl}(H)\\
&=&\left[ \frac{3z}{(z+1)(z-2)}-\frac{1}{2} \right] B^{+}(H).
\end{eqnarray}
Here $E_k^0=\hbar ^2/2m\xi ^2_0(0)$, and $z=n_iJS/M_sH_0$. Notice that
within our model the asymmetry of slopes ratio $B^{-}(H)/B^{+}(H)$
originates from the balance of the exchange $n_iJS$ and localization
induced magnetic $M_sH_0$ energies.

\subsection{Magnetization and the critical temperatures}

Before turning to the comparison of our theoretical findings with the
experimental data, let us discuss the
critical temperatures which control the magnetic ($T_C$) and
carrier localization "metal-insulator" ($T_{MI}$) phase transitions.
According to the adopted model, these two temperatures are defined through
the spontaneous magnetization $M=M_{av}+M_{fl}^{-}$ as follows: $M(T_C)=0$
and $M(T_{MI})=M_0$. Here $M_0\propto H_0$ is the critical magnetization
at which the zero-temperature localization length $\xi _0(H)=\xi _0(0)
(1-H/H_0)^{-1}\propto (1-M/M_0)^{-1}\to \infty$ marking the M-I phase
transition.
According to Section III, the average magnetization reads
$M_{av}(T)\equiv {\cal M}(\eta _0)=M_s\eta _0^{2}(T)$, where $M_s=n_i\mu _B$
is the saturated
magnetization, and the equilibrium order parameter $\eta _0(T)$ is
defined by Eq.(4). Now, for the self-consistency of our approach, we need to
find the fluctuation contributions to the magnetization as well.
Following the lines of the previous Section, we obtain
\begin{equation}
M_{fl}^{-}(T,H)=\frac{Ck_BT^{*}}{4\alpha (H)(T^{*}-T)+4\zeta (H)},
\qquad T\le T^{*}
\end{equation}
and
\begin{equation}
M_{fl}^{+}(T,H)=\frac{Ck_BT^{*}}{2\alpha (H)(T-T^{*})-2\zeta (H)},
\qquad T\ge T^{*}
\end{equation}
As usual, to fix the constant $C$, we assume that $M(T^{*})=M^{+}(T^{*})$,
where $M^{+}=M_{fl}^{+}$ is the magnetization above $T^{*}$. As a result,
we obtain $C=-4M_s\zeta ^{2}/3k_B\beta T^{*}$ which leads to the following
expression for the total magnetization below $T^{*}$
\begin{equation}
M=M_{av}+M_{fl}^{-}=M_s\left(\eta _{0}^{2}-
\frac{\zeta ^2}{3\beta ^2\eta _{0}^{2}} \right),
\end{equation}
with $\zeta$, $\beta $, and $\eta _0$ defined earlier.
Given the above definitions, the two critical temperatures
are related to each other and to the magneto-TEP peak temperature $T^{*}$
within our model as follows
\begin{equation}
T_{MI}=\left(1-\frac{2M_0H_0}{n_eE_k^0-n_iJS}\right)T_C,
\end{equation}
with
\begin{equation}
T_C=\left(1+\frac{yn_iJS}{n_eE_k^0}\right)T^{*}, \qquad y=1-\frac{1}{\sqrt{3}}.
\end{equation}
Let us compare now the obtained theoretical expressions with our 
experimental data on $La_{0.6}Y_{0.1}Ca_{0.3}MnO_3$ (see Fig.2).
By comparing the ratios $(B ^{-}(H)/B ^{+}(H))_{exp}$ and
$(B ^{-}(H)/B ^{+}(H))_{th}$, we obtain $z\simeq 3$ for the slopes
asymmetry parameter leading to $JS=3\mu _BH_0$.
Then, using Eq.(18), $B_{exp}^{+}$, $T^{*}=170K$, and just obtained $z$,
we get $E_k^0/JS=2.5(n_i/n_e)$ which in turn brings about $T_C=195K$
for the Curie temperature (this value falls into the reported range
of the FM transition temperatures for this class of manganites~\cite{5,6,7,8}).
Using this temperature and assuming $S=2$ for an effective Mn spin, we can
estimate the value of the exchange energy $J$ (via the mean-field
expression for the critical field $H_0=3k_BT_C/2S\mu _B$). The result is:
$JS=40meV$, which agrees with other reported estimates of
this parameter.~\cite{11} Besides, from Eq.(23) we immediately get a simple
relationship between the two critical temperatures,
$T_{MI}/T_C=1-4M_0/9M_s$ which allows us to estimate the critical
magnetization $M_0$ (related to the localization magnetic field
$H_0=\mu _0M_0$).
Using $T_{MI}=160K$ (deduced from the GMR data on the
same sample as a peak temperature, see Fig.1), we obtain $M_0=0.4M_s$, in
a good agreement with the localization theory prediction.~\cite{13} Next,
with the
above estimates in mind, Eq.(17) yields $\xi _0=10\AA$ for the
localization length~\cite{5,13} (using a free electron mass $m_e$ for $m$).
Finally, observing that
$JS\simeq k_BT_C\simeq 0.3E_k^0$ we obtain $n_e/n_i=2/3$ for an estimate of
the
free-to-localized carrier number density ratio which leads to the
saturated magnetization $M_s=n_i\mu _B=(3/2)n_e\mu _B$. It is also worth
noting that the found localization energy $E_k^0$ is of the order of
the Fermi energy $E_F$, as expected for manganites.~\cite{11}
To conclude with the estimates, we note that
$\zeta (H)T^{*}/\alpha (H)\simeq 1$ which {\it a posteriori} justifies
the use of the linearized Eq.(13) for the fluctuation region
$|1-T/T^{*}|\ll 1$.
As is seen in Fig.2, this criterion is well met in our case.

In summary, to account for the observed temperature
dependence of the magneto-TEP $\Delta S(T,H)$ in
$La_{0.6}Y_{0.1}Ca_{0.3}MnO_3$,
exhibiting a field-dependent peak at some temperature $T^{*}$ (lying
in-between the charge carrier localization temperature $T_{MI}$ where the
observed negative magnetoresistivity has a minimum, and magnetic transition
temperature $T_C$ which marks the occurence of the spontaneous magnetization),
we adopted the ideas of the localization model and introduced
a free energy functional of Ginzburg-Landau (GL) type describing the phase
transition from paramagnetic (insulator) to ferromagnetic (metal) state
near $T^{*}$. Calculating both average and fluctuation contributions to
the total magnetization and magneto-TEP within the GL theory, we were able
to successfully fit the data and estimate
some important model parameters (including the metal-insulator $T_{MI}$
and magnetic $T_C$ transition temperatures, localization length $\xi _0$,
electron-spin exchange coupling constant $J$, and the free-to-localized
carrier number density ratio $n_e/n_i$), all in a reasonable agreement
with existing microscopic theories.
The Gaussian fluctuations both above and below $T^{*}$ are found to
substantially contribute to the peak value $S_p(H)\equiv \Delta S(T^{*},H)$
of the observed magneto-TEP, amounting to $67\%$ and $33\%$, respectively.

\acknowledgments

We thank J. C. Grenet and R. Cauro (University of Nice-Sophia Antipolis)
for lending us the sample.
Part of this work has been financially supported by the Action de Recherche
Concert\'ees (ARC) 94-99/174. M.A. and A.G. thank CGRI for financial support
through the TOURNESOL program.
S.S. acknowledges the financial support from FNRS (Brussels).

\begin{figure}[htb]
\epsfxsize=8cm
\centerline{\epsffile{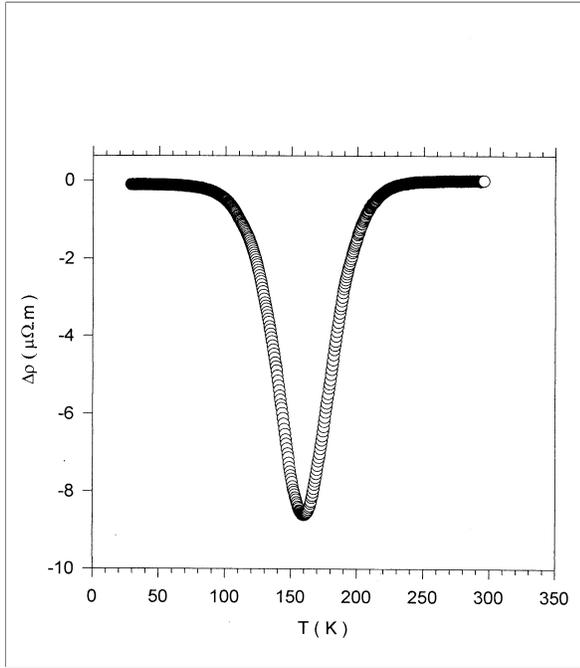} }
\caption{The temperature behavior of the observed magnetoresistivity
in $La_{0.6}Y_{0.1}Ca_{0.3}MnO_3$ at $H=1T$.}
\end{figure}

\begin{figure}[htb]
\epsfxsize=8cm
\centerline{\epsffile{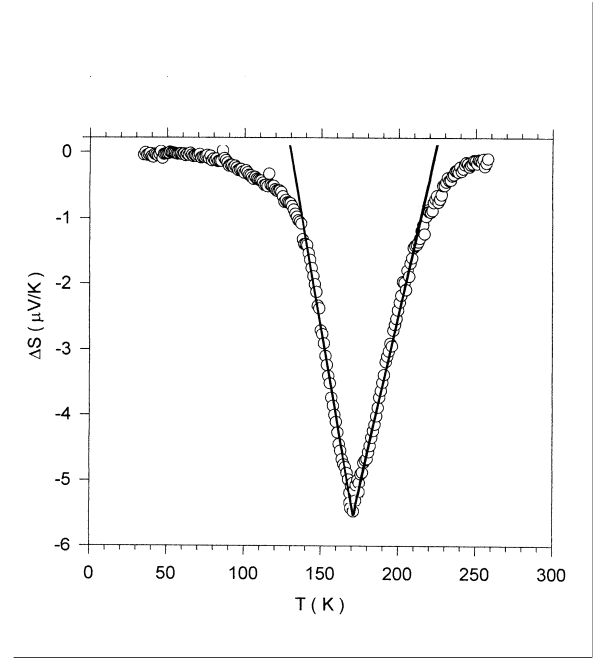} }
\caption{The temperature behavior of the observed magneto-TEP
in manganite $La_{0.6}Y_{0.1}Ca_{0.3}MnO_3$ at $H=1T$.
The best
fit to the data points according to Eq.(1) yields $S_p(H) =
-5.49\pm 0.01 \mu V/K$, $B^{-}(H) = -0.14\pm 0.01 \mu V/K^2$,
and $B^{+}(H) = -0.08\pm 0.01 \mu V/K^2$ for the peak and the slopes,
respectively.}
\end{figure}

\end{document}